# A Systematic Mapping Study on Chatbots in Programming Education

**Marcelino Garcia[3], Renato Garcia[4], Arthur Parizotto[1], Andre Mendes[1]**
**Pedro Valle[4], Ricardo Vilela[5], Renato Balancieri[3]**
**Williamson Silva[2,6]**

[1]Universidade Federal do Pampa - UNIPAMPA (Alegrete), Alegrete, RS, Brasil

[2]PPGES (UNIPAMPA - Alegrete), Alegrete, RS, Brasil

[3]Universidade Estadual de Maringá (UEM), Maringá, PR, Brazil

[4]Instituto de Matemática e Estatística (IME-USP), São Paulo, SP, Brasil

[5]Universidade Estadual de Campinas (FT-Unicamp), Limeira, SP, Brazil

[6]Universidade Federal do Cariri (UFCA), Juazeiro do Norte, CE, Brazil

```
{³pg406476,³rbalancieri}@uem.br
```

```
{¹andremiranda.aluno,¹arthurparizotto.aluno}@unipampa.edu.br
```

```
⁵rfvilela@unicamp.br,{⁴pedrohenriquevalle,⁴renato.s.garcia}@usp.br
```

```
²,⁵williamson.silva@ufca.edu.br
```

***Abstract.*** *Educational chatbots have gained prominence as support tools for teaching programming, particularly in introductory learning contexts. This paper presents a Systematic Mapping Study (SMS) that investigated how such agents have been developed and applied in programming education. From an initial set of 3,216 publications, 54 studies were selected and analyzed based on five research subquestions, addressing chatbot types, programming languages used, educational content covered, interaction models, and application contexts. The results reveal a predominance of chatbots designed for Python instruction, focusing on fundamental programming concepts, and employing a wide variety of pedagogical approaches and technological architectures. In addition to identifying trends and gaps in the literature, this study provides insights to inform the development of new educational tools for programming instruction.*

## 1. Introduction

Programming poses challenges from both didactic and cognitive perspectives, hinder- ing its assimilation by students and its effective mediation by instructors [Robins 2019, Luxton-Reilly 2016]. As a result, introductory programming courses have historically shown high rates of failure [Alves et al. 2019] and dropout [Penney et al. 2023], under- scoring the need for pedagogical strategies that combine individualized support, instructional clarity, and continuous engagement. These difficulties go beyond technical content, being closely related to how knowledge is presented, contextualized, and internalized by learners [Guo 2018, López-Pernas et al. 2019]. According to Cognitive Load The- ory [Sweller et al. 2011], effective learning environments must reduce unnecessary de-mands on working memory by offering structured and context-sensitive instruction tailored to learners' proficiency levels. In this sense, instructional personalization — that is, adjusting the level of guidance to learners' cognitive maturity — is a key factor in ensuring accessibility and meaningful learning [Van Merrienboer e Sweller 2005, Renkl 2014].

This scenario has driven the development of solutions that integrate technological affordances with sound pedagogical principles. Among these, chatbots have emerged as promising tools for continuous learning mediation, offering imme- diate feedback and fostering learner autonomy [Ruan et al. 2019, Clarizia et al. 2018, Smutny e Schreiberova 2020]. These intelligent applications are capable of interacting with learners through natural language or pre-structured dialogues [Mageira et al. 2022], and can be integrated into virtual learning environments or personalized educational sys- tems, enabling various forms of automated and adaptive tutoring digital scaffolding mech- anisms [Hobert 2019, Carreira et al. 2022, Kasinathan et al. 2018]. These tools extend learners' exposure time to instructional content, creating opportunities for deliberate prac- tice [Ericsson et al. 1993], in which students experiment, test hypotheses, refine their understanding, and internalize concepts actively and reflectively.

Despite the growing number of initiatives involving the use of chatbots in Education, the literature still lacks comprehensive systematizations that consolidate the existing body of knowledge, particularly regarding their adoption in undergraduate programming education. The dispersed nature of current publications across various sources makes it difficult to identify methodological patterns, pedagogical foundations, and application contexts. Organizing and analyzing such evidence is essential not only to support more informed decision-making but also to foster the development of more effective educational solutions that align with the instructional needs of both students and educators. Given this context, in this paper, we present a Systematic Mapping Study (SMS) car- ried out to investigate how chatbots have been adopted in the teaching and learning of programming in undergraduate courses. SMS is a research method designed to categorize and synthesize scientific evidence in a rigorous, transparent, and reproducible manner, enabling the identification of gaps, emerging trends, and directions for future research [Kitchenham e Charters 2007, Petersen et al. 2015].

Based on the analysis of 2,497 retrieved studies, we selected 34 primary studies through the rigorous application of predefined inclusion and exclusion criteria. Our findings offer a comprehensive, structured, and reproducible overview of the state of the art in chatbot-based programming education, providing valuable insights from both technical and pedagogical perspectives. The main contributions of this study are as follows: (i) a catalog of chatbots proposed for programming education; (ii) the identification of the programming contents and concepts addressed by these solutions; (iii) the analysis of interaction strategies used to mediate the learning process; and (iv) the mapping of programming languages employed in educational interactions supported by chatbots.

## 2. Research Method

In this study, we conducted a Systematic Mapping Study (SMS) following the guidelines proposed by Kitchenham et al. (2007). Our goal is to investigate the chatbots described in the literature that support the teaching of programming in undergraduate courses. We detailed the procedures adopted in this study in the following subsections.

## 2.1. Research Questions

The following research question guided our study: **How have chatbots been employed as pedagogical tools in undergraduate programming education?** We defined a set of Research subquestions (SQs) to guide the data extraction and analysis process. Table 1 presents the SQs along with their respective motivations.

**Table 1. Research subquestions and their motivations.**

| Research Subquestions | Motivation |
| --- | --- |
| **SQ1.** Which chatbots have been proposed to support programming learning? | To identify and systematize chatbots described in the scientific literature that focus on programming education, providing a comprehensive overview of the solutions already developed or evaluated in educational contexts. |
| **SQ2.** What programming concepts are addressed by educational chatbots | To understand which programming concepts, skills, or domains are most frequently targeted through chatbot-based learning, highlighting trends and areas of concentrated use. |
| **SQ3.** Which programming languages are employed in chatbot-mediated learning interactions? | To investigate the diversity and prevalence of programming languages used in educational chatbot approaches, and to map their suitability across different levels of complexity and learner profiles. |
| **SQ4.** What interaction strategies are used by educational chatbots? | To identify the communication mechanisms employed by chatbots (Pattern-based models ( menu/script-based interactions), Retrieval-based models, Generative models). |

## 2.2. Digital Databases

We selected four digital libraries to conduct our SMS: ACM Digital Library (ACM), Engineering Village (Compendex), IEEE Xplore Digital Library (IEEE) e Scopus. We chose these databases for their recognized relevance in Computing Education and their broad coverage of high-quality scientific publications.

## 2.3. Search Strategy

We adopted a refinement process similar to the one proposed by Zhang et al. (2010), which included the analysis of previously selected studies and the use of control papers to validate the effectiveness of the formulated search strings. We applied the search string presented below for automated retrieval in the selected digital libraries, combining alternative spellings and synonyms using the OR boolean operator. Additionally, we used the AND operator to combine the three core concepts of interest.

> ("chatbot" **OR** "chatterbot" **OR** "artificial conversational entity" **OR** "chatbots" **OR** "mobile chatbots" **OR** "conversational agent*" **OR** "talkbot*" **OR** "talk bot*" **OR** "conversational interface" **OR** "conversational system" **OR** "dialogue system") **AND** ("programming*" **OR** "program" **OR** "CS2" **OR** "CS1" **OR** "computer programming" **OR** "introductory computer" **OR** "introductory programming" **OR** "novice programming" **OR** "coding education" **OR** "introductory computer science") **AND** ("student" **OR** "learning" **OR** "course" **OR** "teaching" **OR** "training")

## 2.4. Inclusion and Exclusion Criteria

We defined the Inclusion Criteria (IC) and Exclusion Criteria (EC), as outlined in the guidelines by Kuhrmann et al. (2017). Table 2 presents the criteria adopted in this SMS.

Table 2. Criteria for inclusion or exclusion of studies.

| ID | Description |
|---|---|
| IC1 | Studies that discuss the use of chatbots to support programming learning in undergraduate education. |
| IC2 | Studies that report empirical evidence (e.g., experiments, evaluations) involving chatbot use in undergraduate programming education. |
| IC3 | Studies that describe chatbots as learning support tools for programming in undergraduate contexts. |
| EC1 | Studies that address educational chatbots outside the scope of programming education. |
| EC2 | Duplicate studies (e.g., versions published in different venues or dates); In this case, we considered only the most complete and latest version. |
| EC3 | Studies that are not written in English. |
| EC4 | The following types of publication: books, doctoral theses, master's dissertations, patents, tutorials, workshop proposals, or posters. |
| EC5 | The study's full text is not available for download. |

## 2.5. Establishment of the Selection Process

We adopted the search string presented in Subsection 2.3 to retrieve the candidate studies for analysis in this SMS, conducted in April 2025. We carried out the selection process iteratively and incrementally, in two main stages. In the **first stage**, we screened studies based on their titles and abstracts, applying the inclusion and exclusion criteria described in Subsection 2.4. Two researchers jointly discussed decisions. In cases of doubt or lack of consensus, we retained the study for the next stage of analysis. In the **second stage**, we performed a full-text review of the previously selected studies, systematically applying the predefined selection criteria. This stage aimed to ensure a more accurate assessment of each study's relevance. We reviewed and discussed all results collaboratively, and we resolved any disagreements through consensus among the researchers.

Figure 1 summarizes the complete data selection and extraction process. Initially, we found 3,216 publications in the digital libraries: 112 from the ACM Digital Library, 855 from Engineering Village, 984 from IEEE Xplore, and 1,265 from Scopus. Some articles were indexed in more than one database but were counted only once, following a predefined priority order: ACM, then Engineering Village, IEEE, and finally Scopus. We accessed all potential studies through our institutional network. When full-text access was not available, we contacted the authors directly to request preprint versions. After applying the inclusion and exclusion criteria, we selected a final set of 54 studies to be included in this SMS.

## 2.6. Data Extraction Strategy

After thoroughly reading each study, we systematically extracted relevant information to answer each research subquestion presented in Table 1 Two additional authors subsequently reviewed all extracted data to ensure the consistency and accuracy of the collected information. Any disagreements were discussed among the researchers until a consensus was reached.

## 3. Results

Table 3 presents the complete list of selected studies.

The selected studies were published between 2003 and 2025. From the temporal perspective, the earliest relevant study was published in 2003, followed by a ten-year gap. A new study emerged in 2013, but once again, there was a hiatus until 2018, when we identified three relevant publications. After a slight decline in 2020, we observed a steady increase in the following years, with four studies published in 2021 and five in 2022. This increase may be associated with significant advancements in the field of Natural Language Processing (NLP), particularly following the introduction of the Transformer architecture in 2017, which enabled the development of more sophisticated chatbots with enhanced contextual understanding [Bran e Schwaller 2024]. In addition, the COVID-19 pandemic created an urgent need to restructure teaching and learning models, demanding accessible and interactive educational solutions. These factors, combined, have contributed to the growing interest among educators in adopting chatbots for programming education.

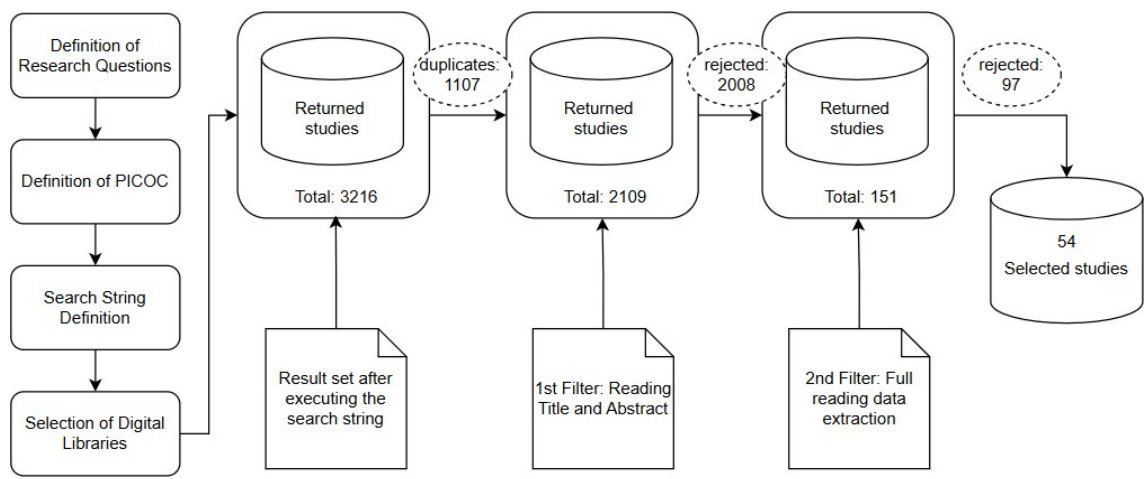

Figure 1. Results of systematic mapping filters.

Table 3. Selected studies.

| ID | Reference | ID | Reference | ID | Referece |
|---|---|---|---|---|---|
| S01 | [Coronado et al. 2018] | S19 | [Kumar et al. 2024] | S37 | [Wijaya e Purwarianti 2024] |
| S02 | [Verleger e Pembridge 2018] | S20 | [Gupta et al. 2025b] | S38 | [Vintila 2024] |
| S03 | [Lane e VanLehn 2003] | S21 | [Gupta et al. 2025a] | S39 | [Liu et al. 2024] |
| S04 | [Nguyen et al. 2022] | S22 | [Frankford et al. 2024] | S40 | [Lin 2022] |
| S05 | [Ardimansyah e Widianto 2021] | S23 | [Xue et al. 2024] | S41 | [Scholl et al. 2025] |
| S06 | [Chen et al. 2021] | S24 | [Vadaparty et al. 2025] | S42 | [da Silva et al. 2024] |
| S07 | [Bilgin e Yavuz 2022] | S25 | [Troussas et al. 2024] | S43 | [Kosar et al. 2024] |
| S08 | [Ismail e Ade-Ibijola 2019] | S26 | [Lee et al. 2024] | S44 | [Modran et al. 2024] |
| S09 | [Carreira et al. 2022] | S27 | [Wei et al. 2023] | S45 | [Zabala e Narman 2024] |
| S10 | [Okonkwo e Ade-Ibijola 2020] | S28 | [Kiesler et al. 2023] | S46 | [Haindl e Weinberger 2024a] |
| S11 | [Okonkwo e Ade-Ibijola 2022] | S29 | [Cubillos et al. 2025] | S47 | [Lepp e Kaimre 2025] |
| S12 | [Winkler et al. 2020] | S30 | [Pirzado et al. 2024] | S48 | [Callejo et al. 2024] |
| S13 | [Hobert 2019] | S31 | [Arteaga Garcia et al. 2024] | S49 | [Fernandez et al. 2024] |
| S14 | [Kuo e Chen 2023] | S32 | [Palahan 2025] | S50 | [Farah et al. 2023] |
| S15 | [Kasinathan et al. 2018] | S33 | [Bobadilla et al. 2023] | S51 | [Tayeb et al. 2024] |
| S16 | [Hamzah et al. 2021] | S34 | [Gabriella et al. 2024] | S52 | [Haindl e Weinberger 2024b] |
| S17 | [Mitchell et al. 2013] | S35 | [Xiao et al. 2024] | S53 | [Prather et al. 2024] |
| S18 | [Groothuijsen et al. 2024] | S36 | [Akçapınar e Sidan 2024] | S54 | [Andersen-Kiel e Linos 2024] |

We present the results for each of the outlined SQs below.

## 3.1. SQ1. Which chatbots have been proposed to support programming learning?

From the 54 selected studies, we identified and characterized 36 distinct chatbots designed to support programming learning.

We organized the identified chatbots according to the predominant pedagogical strategies adopted in their design. This categorization allowed us to recognize instructional patterns that transcend technological mediation, highlighting distinct modes of interaction with knowledge, learners, and the educational context.

The definition of these pedagogical categories was informed by theoretical guidelines from [Renkl 2014, Keuning et al. 2018, Okonkwo e Ade-Ibijola 2021] and derived inductively through a systematic reading of the studies. We grouped the chatbots into the following categories:

- **Content-based direct instruction.** This category includes chatbots designed to deliver structured explanations, definitions, and direct instructional support related to programming syntax, commands, and foundational concepts. These agents typically follow a transmissive pedagogical approach, focusing on the clear presentation of knowledge through predefined responses or rule-based logic. They are especially suited for novice learners, helping to reinforce basic understanding and support early-stage learning. Examples include: RPL Course (S5), Lecturer's Apprentice (S8), Coding Tutor (S13), JavaTutor (S17), Duke (S1), Web Programming (S16), WeLast (S16), Intelligence Concentration Framework (S39), Chatbot Script (S40), and inumerous ChatGPT-based implementations (S19, S34, S41, S42, S43, S44, S48, S49, S51, S52, S53, S54).
- **Reflective and metacognitive tutoring.** This category comprises chatbots designed to promote self-regulated learning by incorporating pedagogical strategies such as Socratic questioning, emotional scaffolding, and mediation of reasoning processes. Rather than merely transmitting content, these agents aim to stimulate learners' critical thinking, enhance metacognitive awareness, and reduce cognitive load through dialogic interaction. They often adapt their responses to encourage reflection, support the development of learning strategies, and deepen conceptual understanding. Examples include: Edubot (S2), Pseudocode Tutor (S3), Anonymous Chatbot (S4), Disha AI (S21), ChatGPT-based implementations (S18, S27, S28), Gemini 1.5 (S29), Newton (S31), PythonPal (S32), QuickTA (S35), and Chatbot Avert (S38).
- **Automated feedback on programming tasks.** This category encompasses chatbots designed to provide immediate and automated feedback on programming exercises. These agents analyze submitted code, identify syntactic or logical errors, and offer tailored suggestions for improvement. Their primary goal is to support learners in debugging and refining their code autonomously, thereby reinforcing concepts through iterative practice and reducing dependence on instructor intervention. Many of these chatbots integrate AI-powered techniques—especially generative models—to enhance the clarity, adaptiveness, and pedagogical effectiveness of their responses. Examples include: Copilot (S24), ChatGPT-4 (S26), SOBO (S33), AI Programming Assistant (S36), ChatGPT-based implementations (S18, S25, S45, S46, S47), and Rule-Based Chatbot (S50).
- **Contextual and multimodal tutoring.** This category includes chatbots that enrich the instructional experience by incorporating multimodal resources—such as videos, audio narration, and annotated materials—into the learning process. These agents are typically deployed in synchronous or blended learning environ- ments, where they provide real-time or near-real-time guidance aligned with the presented content. By leveraging diverse media formats, these chatbots aim to enhance engagement, contextual understanding, and retention, particularly for visual and auditory learners. Examples include: Sara (S12), Anonymous Chatbot (S14), TicTad (S15), and Intelligent Tutoring QA System (S37).

- **Autonomous instruction and self-study support.** This category comprises chatbots designed to support individual and self-paced learning, often embedded directly within coding environments or functioning as on-demand tutors. These agents provide personalized guidance without requiring instructor mediation, offering features such as intelligent feedback, scheduled sessions, interactive exercises, and support for advanced topics. Their goal is to foster learner autonomy, reinforce continuous engagement, and promote knowledge retention in both introductory and advanced programming contexts. Examples include: PyNar (S7), Pyo (S9), Python-Bot (S10), Revision-Bot (S11), AI Tutor (S22), Newton (S31), ChatGPT (S23, S52).

- **Practical assistance in domain-specific contexts.** This category includes chatbots developed to support programming in applied or specialized domains, such as embedded systems, hardware integration, or large-scale educational settings. These agents are tailored to context-specific needs, providing targeted guidance, technical assistance, or organizational support beyond traditional instructional content. Their role is to bridge the gap between programming knowledge and real-world applications, enhancing the authenticity and relevance of learning experiences. Examples include: GSI Agent (S1), FritzBot (S6), and Anonymous Chatbot (S20).

The diversity of pedagogical strategies identified — ranging from direct instruction to adaptive and metacognitive support — reflects the growing maturity of chatbot-based learning approaches and the initial incorporation of theoretical foundations into agent design. However, we observed that many studies still emphasize technological innovation at the expense of explicit instructional grounding, reinforcing the need for evidence-based pedagogical models in the design of educational chatbots. This analy- sis provides valuable insights for understanding the current state of the field and guiding future research that combines pedagogical rigor with technical sophistication in the development of conversational agents for programming education.

### 3.2. SQ2. What programming concepts are addressed by educational chatbots?

The analysis of the selected studies revealed that educational chatbots have primarily focused on supporting foundational programming concepts. Among the 54 studies analyzed, we identified six thematic categories based on the instructional content addressed by the chatbots:

- **Language-Specific Programming** (S1–S5, S7–S13, S15–S17, S25–S28, S31, S32, S34–S42, S44, S45, S48, S50, S51, S53, S54): The majority of chatbots were designed to teach introductory programming concepts—such as variables, control structures, and basic syntax—through specific programming languages, especially Python, Java, and C-family languages. This prevalence reflects a pedagogical emphasis on early-stage skill development, where syntactic fluency and basic logic are critical. This prevalence reflects a pedagogical emphasis on early-stage skill development, where syntactic fluency and basic logic are critical. Examples include: Python-Bot (S10), JavaTutor (S17), Coding Tutor (S13), and WebProgramming (S16).

- **General Programming Concepts** (S18, S20–S22, S24, S30, S33, S46, S47, S49, and S52): Several chatbots (e.g., Disha AI – S21, ChatGPT – S18, S24) addressed core programming logic, problem-solving skills, and algorithmic reasoning without being tied to a specific language. These tools are especially useful for scaffolding abstract thinking and conceptual understanding, often employing natural language explanations or pseudocode to reduce cognitive load for beginners.
- **Object-Oriented Programming (OOP)** (S3, S23, S43, S46, S47, S49, S52–S54): A subset of chatbots focused on learning OOP principles, such as classes, inheritance, and encapsulation. These include S23, S43, and S52, often leveraging ChatGPT to support student understanding of advanced concepts in Java or C++.
- **Data Structures** (S14 and S29): Two studies used chatbots to support learning in the context of data structures learning, suggesting a potential yet underexplored application area for educational chatbots.
- **Web Programming** (S19): One study focused on teaching web development concepts through interactions with ChatGPT and other MLLMs, highlighting opportunities for applying chatbots in frontend/backend education.
- **Physical Computing** (S6): FritzBot was the only chatbot aimed at physical computing and embedded systems, supporting students in programming with Arduino and prototyping.

An analysis of the included studies revealed that some chatbots addressed multiple programming content categories simultaneously (S4, S46, S47, and S49), illustrating the potential for broader instructional scope. For instance, Study S4 was classified under both Object-Oriented Programming (OOP) and Language-Specific Programming, as it supported the teaching of basic C++ syntax and structures alongside core OOP concepts such as classes and inheritance. This overlap highlights the possibility of designing educational chatbots capable of supporting learners through different stages of programming knowledge development, from foundational constructs to more abstract paradigms. It also reinforces the need for flexible classification frameworks that reflect the multidimensional nature of programming instruction.

Our results indicate a predominance of chatbots aimed at introductory programming education, reflecting a strong interest in supporting students at the beginner level. However, the presence of chatbots designed for more advanced topics—such as data structures, object-oriented programming, and web development—reveals a trend toward diversification and pedagogical maturity. Furthermore, the incorporation of features related to emotional support and institutional mediation points to an ongoing movement toward personalization, scope expansion, and multifunctionality in the use of educational chatbots.

### 3.3. SQ3. Which programming languages have been taught using educational chatbots?

Our analysis of the selected studies revealed a diversity of programming languages employed by educational chatbots designed to support programming education.

**Python** emerged as the most frequently used language (S7, S9–S12, S18–S21, S24, S26, S31, S32, S35, S36, S39, S40, S44, S45, S48, S49, and S54). This prevalence reflects Python's growing popularity in both academic and professional contexts, particularly in introductory programming courses. We also identified chatbots that support other commonly adopted languages, including **Java** (S1, S13, S17, S21, S23, S25, S28, S33, S39, S46, S47, S52, and S54), **C** (S18, S22, S27, S29, S42, S44), **C++** (S4, S21, S43, S53, S54), **C#** (S15, S19, S36), **Matlab** (S2), **JavaScript** (S19), **TypeScript** (S19), **PHP** (S16) and **Arduino** (S6).

In a few cases, **pseudocode** was adopted as an instructional language (S3, S8), serving as a cognitive educational tool between algorithmic thinking and formal coding syntax. Some chatbots focused exclusively on conceptual explanations or textual definitions without relying on any specific programming language (e.g., S5 for natural language explanations and S14 for data structures). Finally, some studies did not explicitly report the programming language employed in the chatbot's design or usage (S30, S34, S37, S38, S41, S49–S51), which may reflect either a focus on pedagogical strategies over technical implementation or an intentional emphasis on language-agnostic instruction.

These findings underscore the diversity of instructional strategies in chatbot-mediated programming education—from language-specific instruction aligned with industry practices to abstract, conceptual approaches designed to reduce cognitive load and support novice learners in building foundational understanding.

### 3.4. SQ4. What interaction strategies are used in the educational chatbots?

The interaction between students and educational chatbots plays an important role in the pedagogical effectiveness of these agents. According to the taxonomy proposed by Hien et al. (2018), chatbots can be categorized into three main types of interaction models:

- **Pattern-based models** rely on fixed input–output rules, typically structured as predefined question–answer pairs. While computationally simpler, these systems provide predictable and constrained interactions, making them suitable for domains with well-defined scopes. Examples include: Duke (S1), Pseudocode Tutor (S3), RPL Course (S5), Lecturer's Apprentice (S8), Anonymous Chatbot (S14), TicTad (S15), Newton (S31), SOBO (S33), Chatbot Script (S40), and Rule-based Chatbot (S50).
- **Retrieval-based models** operate by querying structured repositories or APIs, and returning responses based on semantic similarity or contextual matching. These models enhance interactional flexibility but depend heavily on the quality and coverage of the knowledge base. Examples include: EduBot (S2), Anonymous Chatbot (S4), PyNar (S7), Pyo (S9), Python-Bot (S10), Revision-Bot (S11), Coding Tutor (S13), JavaTutor (S17), SOBO (S33), as well as more recent studies like the Intelligent Tutoring QA System (S37).
- **Generative models** leverage machine learning techniques— particularly LLMs — to produce dynamic, context-aware responses. These agents demonstrate adaptability to natural language and conversational complexity, though they raise concerns about pedagogical coherence and content control. Examples include: FritzBot (S6), Sara (S12), WeLast (S16), WebProgramming (S16), ChatGPT (S18, S19, S23–S28, S34, S41–S48, S49, S51–S54), Anonymous Chatbot (S20), Disha AI (S21), AI Tutor (S22), GitHub Copilot (S24, S53), Google Gemini 1.5 (S29, S49), Newton (S31), PythonPal (S32), QuickTA (S35), AI Programming Assis-tant (S36), Chatbot AVERT (S38), Intelligence Concentration Framework (S39), and Bing (S49).

In addition to these categories, **hybrid architectures** have emerged as a promising alternative, combining multiple approaches to balance robustness and adaptability. For instance, SOBO (S33) integrates rule-based patterns with information retrieval mechanisms, while S27 stands out for incorporating all three models—pattern-based, retrieval-based, and generative—demonstrating the complementary potential of predefined structures, knowledge queries, and contextualized language generation. One study (S30) did not provide sufficient detail to determine the specific interaction model employed by the chatbots. Our data analysis reveals a progressive shift from pattern- and retrieval-based architectures toward generative models, driven by the increasing availability and maturity of LLMs. This trend reflects not only technological advances in AI but also a pedagogical paradigm shift—from traditional instructional delivery to more responsive, personalized, and learner-sensitive experiences. Nevertheless, the adoption of generative models does not negate the relevance of hybrid approaches. These can be particularly effective in contexts requiring both stability and adaptability, such as large-scale introductory programming courses or platforms that combine technical content with emotional or motivational support. In this light, the selection of an interaction model becomes a strategic decision that should account not only for computational resources, but also for instructional goals and the cognitive profiles of the learners being served.

## 4. Discussion

The analysis of our sub-research questions revealed a comprehensive landscape regarding the use of chatbots in programming education, highlighting both technological advancements and pedagogical challenges. In this section, we discuss our findings in light of existing literature, aiming to identify recurring patterns, tensions, and emerging opportunities. By integrating theoretical and practical perspectives, we seek to understand how instructional strategies, targeted content, supported programming languages, and interaction models adopted by chatbots contribute to—or hinder—the development of more effective, personalized, and scalable learning experiences.

### 4.1. Functional Diversity and Maturation of Educational Chatbots

Our chronological analysis of the selected studies (SQ1) reveals a clear evolution in the design of educational chatbots, with noticeable acceleration starting around 2018. Initially limited to rule-based and narrowly scoped applications, contemporary chatbots have begun to incorporate more sophisticated approaches, expanding their roles beyond teaching basic syntax or commands. The presence of chatbots with multiple functionalities—such as emotional support (Lecturer's Apprentice – S8), metacognitive tutoring (Disha AI – S21), and self-paced learning support with code analysis (Newton – S31; PythonPal – S32)—suggests a shift toward personalized learning experiences, positioning chatbots as key components of hybrid and adaptive educational ecosystems. This trend reinforces a sociotechnical perspective of technology-mediated education, where chatbots are not merely content transmitters, but also facilitators of cognitive, emotional, and social relationships between learners and knowledge. As noted by Keuning et al. (2018) , the potential of chatbots lies in their ability to engage with the authentic challenges of pedagogical practice—a priority increasingly reflected in recent studies.

## 4.2. Prevalence of Introductory Content and Gaps in Advanced Topics

Our findings for SQ2 reveal a predominance of chatbots designed to teach introductory programming content, typically structured around specific programming languages. This emphasis reflects a legitimate concern within the educational community to support learners during their initial contact with programming—a stage often marked by high cognitive load, high dropout rates, and difficulties in understanding control flow, syntax, and debugging. However, the scarcity of chatbots targeting more advanced topics—such as data structures, object-oriented programming, or web development—exposes a relevant gap in the field. From an instructional perspective, this limitation reduces the potential of chatbots to serve as mediators throughout the full spectrum of a student's learning journey. Addressing this issue involves not only expanding content coverage but also adapting pedagogical strategies by incorporating principles such as fading, example variation, and self-explanation prompts [Renkl 2014], which are essential for fostering the acquisition of more complex skills.

## 4.3. Adoption of Accessible Languages and Cognitive Transition Strategies

The analysis of SQ3 indicates a substantial prevalence of programming languages such as Python and Java in educational chatbots, which aligns with their widespread use in introductory curricula and professional settings. These choices are not only techni- cally sound but also pedagogically informed, as these languages offer a broad commu- nity support and extensive learning resources that facilitate self-directed learning and re- duce initial barriers. On the other hand, the use of pseudocode in agents such as Pseu- docode Tutor (S3) and Lecturer's Apprentice (S8) demonstrates a conscious adoption of cognitive transition strategies, particularly in contexts that emphasize conceptual under- standing before implementation. This approach is supported by Cognitive Load Theory [Sweller et al. 1998, Sweller et al. 2011], which highlights the importance of minimizing extraneous load in the early stages of learning, and by the principles of progressive scaf- folding that underpin Example-Based Learning. The observed diversity of languages and representations also reflects the need to accommodate student heterogeneity, acknowledg- ing different learning styles, paces, and prior knowledge.

## 4.4. Transition to Generative Models and Their Educational Implications

Our findings for SQ4 suggest a paradigm shift in how chatbots interact with students, moving from rigid pattern-based models to more fluid and adaptive generative models, powered by LLMs such as GPT-3.5, GPT-4, and Gemini. This evolution marks a signif- icant milestone in the integration of AI and education, enabling agents to dynamically respond to learners' needs, questions, and communication styles in real time. However, this technical advancement also introduces new pedagogical challenges. Key concerns in- clude the risk of incorrect or pedagogically unsound responses, the opacity of generative models, and the difficulty of ensuring alignment with instructional goals. As highlighted by Jury et al. (2024) and Pirzado et al. (2024), the mere integration of LLMs does not au- tomatically translate into effective educational outcomes. The conversational architecture of chatbots must be designed around explicit pedagogical principles, such as personalized feedback, reflective reasoning, and the promotion of learner autonomy. In this regard, hybrid solutions—which combine generative models with information retrieval mecha- nisms and predefined instructional patterns—emerge as promising alternatives, offering a balance between pedagogical control and AI responsiveness.

## 4.5. Contemporary Implications and Emerging Challenges

Beyond the direct evidence provided by this SMS, the current literature reinforces the notion that educational chatbots are not merely technical tools, but rather social and cognitive technologies. Their impact extends beyond operational efficiency, influencing student engagement, autonomy, confidence, and knowledge construction. The widespread use of Python is not solely a result of its syntactic simplicity, but also of its strong integration with NLP libraries and educational platforms. Examples such as PythonPal and Python-Bot illustrate how technical integration can enhance instructional support. However, several latent challenges remain. The continuous use of chatbots—if not guided by intentional pedagogical design—may undermine the development of critical thinking skills. The replacement of human mediation by automated agents, when poorly planned, tends to reduce opportunities for argumentation, idea exchange, and collaborative learning. Therefore, the future of educational chatbots will depend on our ability to combine technological innovation with robust pedagogical principles, such as personalization, formative feedback, and the fostering of reflective reasoning.

## 5. Final Considerations

Our SMS investigated how chatbots have been developed and applied to support the teaching and learning of programming in undergraduate contexts. In response to the research question presented in Section 2.1, we found that chatbots have been predominantly employed to support the learning of introductory programming concepts, especially using languages such as Python, Java, and C++. These agents typically serve as automated tutors, offering on-demand explanations, immediate feedback, problem-solving support, code suggestions, and self-study assistance. Most chatbots analyzed were designed to interact responsively with students, ranging from pattern-based systems to, more recently, generative natural language models powered by LLMs. Despite this evolution, we observed a predominance of systems lacking grounding in learning theories, with minimal explicit instructional foundations. Furthermore, there is a clear concentration on basic programming topics, with limited focus on more advanced areas such as data structures, object-oriented programming, or web development.

This study contributes to the field by organizing and characterizing underexplored gaps—namely, the scarcity of agents designed for advanced programming topics and the absence of pedagogical guidelines for chatbot design in computer science education. By compiling and analyzing these findings, we provide valuable insights for the development of future educational tools and research agendas that focus on the effective integration of chatbots, learning theories, and pedagogical practices. We hope our results inspire the design of agents that go beyond automated response systems, serving instead as dialogic mediators of the learning process—sensitive to learners' cognitive needs and aligned with contemporary theories in computing education. For future work, we recommend systematically investigating the impact of different types of chatbots on student learning, considering variables such as academic performance, engagement, motivation, and the development of cognitive skills. Rather than merely expanding chatbot presence in classrooms, we emphasize the importance of ensuring that their use aligns with principles of equity, instructional clarity, and the promotion of learner autonomy.

# References

Akçapınar, G. e Sidan, E. (2024). Ai chatbots in programming education: guiding success or encouraging plagiarism. *Discover Artificial Intelligence*, 4(87):1–18.

Alves, G., Rebouças, A., e Scaico, P. (2019). Coding dojo como prática de aprendizagem colaborativa para apoiar o ensino introdutório de programação: Um estudo de caso. Em *Anais do XXVII Workshop sobre Educação em Computação*, páginas 276–290. SBC.

Andersen-Kiel, N. e Linos, P. (2024). Using chatgpt in undergraduate computer science and software engineering courses: A students' perspective. Em *Proceedings of the 2024 IEEE Frontiers in Education Conference (FIE)*, páginas 1–9. IEEE.

Ardimansyah, M. e Widianto, M. (2021). Development of online learning media based on telegram chatbot (case studies: Programming courses). Em *Journal of Physics: Conference Series*, volume 1987, página 012006. IOP Publishing.

Arteaga Garcia, E. J., Nicolaci Pimentel, J. F., Feng, Z., Gerosa, M., Steinmacher, I., e Sarma, A. (2024). How to support ml end-user programmers through a conversational agent. Em *Proceedings of the 46th IEEE/ACM International Conference on Software Engineering*, páginas 1–12.

Bilgin, T. T. e Yavuz, E. (2022). Integrating a dialogue tree based turkish chatbot into an open source python coding editor. Em *2022 3rd International Informatics and Software Engineering Conference (IISEC)*, páginas 1–5. IEEE.

Bobadilla, S., Glassey, R., Bergel, A., e Monperrus, M. (2023). Sobo: A feedback bot to nudge code quality in programming courses. *IEEE Software*, 41(2):68–76.

Bran, A. M. e Schwaller, P. (2024). Transformers and large language models for chemistry and drug discovery. Em *Drug Development Supported by Informatics*, páginas 143–163. Springer.

Callejo, P., Alario-Hoyos, C., e Delgado-Kloos, C. (2024). Evaluating chatgpt impact on the programming learning outcomes of students in a big data course. *International Journal of Engineering Education*, 40(4):863–872.

Carreira, G., Silva, L., Mendes, A. J., e Oliveira, H. G. (2022). Pyo, a chatbot assistant for introductory programming students. Em *2022 International Symposium on Computers in Education (SIIE)*, páginas 1–6. IEEE.

Chen, T., Xu, L., e Zhu, K. (2021). Fritzbot: A data-driven conversational agent for physical-computing system design. *International Journal of Human-Computer Studies*, 155:102699.

Clarizia, F., Colace, F., Lombardi, M., Pascale, F., e Santaniello, D. (2018). Chatbot: An education support system for student. Em *International Symposium on Cyberspace Safety and Security*, páginas 291–302. Springer.

Coronado, M., Iglesias, C. A., Carrera, Á., e Mardomingo, A. (2018). A cognitive assistant for learning java featuring social dialogue. *International Journal of Human-Computer Studies*, 117:55–67.

Cubillos, C., Mellado, R., Cabrera-Paniagua, D., e Urra, E. (2025). Generative artificial intelligence in computer programming: Does it enhance learning, motivation, and the learning environment? *IEEE Access*.

da Silva, C. A. G., Ramos, F. N., de Moraes, R. V., e dos Santos, E. L. (2024). Chatgpt: Challenges and benefits in software programming for higher education. *Sustainability*, 16(3):1245.

Ericsson, K. A., Krampe, R. T., e Tesch-Römer, C. (1993). The role of deliberate practice in the acquisition of expert performance. *Psychological review*, 100(3):363.

Farah, J. C., Spaenlehauer, B., Ingram, S., Purohit, A. K., Holzer, A., e Gillet, D. (2023). Harnessing rule-based chatbots to support teaching python programming best practices. Em *Proceedings of the International Conference on Educational Technology*. Springer.

Fernandez, C., Sánchez-Soto, E., Aguilar Cisnero, J., e Juárez-Ramírez, R. (2024). Exploring the frontier of software engineering education with chatbots. *Programming and Computer Software*, 50(8):796–815.

Frankford, E., Sauerwein, C., Bassner, P., Krusche, S., e Breu, R. (2024). Ai-tutoring in software engineering education. Em *Proceedings of the 46th International Conference on Software Engineering: Software Engineering Education and Training*, páginas 309–319.

Gabriella, A., Gui, A., e Chanda, R. C. (2024). The use of chatbot and its impact on academic achievement. Em *2024 IEEE Symposium on Industrial Electronics & Applications (ISIEA)*, páginas 1–6. IEEE.

Groothuijsen, S., van den Beemt, A., Remmers, J. C., e van Meeuwen, L. W. (2024). Ai chatbots in programming education: students' use in a scientific computing course and consequences for learning. *Computers and Education: Artificial Intelligence*, 7:100290.

Guo, P. J. (2018). Non-native english speakers learning computer programming: Barriers, desires, and design opportunities. Em *Proceedings of the 2018 CHI conference on human factors in computing systems*, páginas 1–14.

Gupta, R., Goyal, H., Kumar, D., Mehra, A., Sharma, S., Mittal, K., e Challa, J. S. (2025a). Sakshm ai: Advancing ai-assisted coding education for engineering stu- dents in india through socratic tutoring and comprehensive feedback. *arXiv preprint arXiv:2503.12479*.

Gupta, R. D., Hosain, M. T., Mridha, M., e Ahmed, S. U. (2025b). Multimodal programming in computer science with interactive assistance powered by large language model. Em *International Conference on Human-Computer Interaction*, páginas 59–69. Springer.

Haindl, P. e Weinberger, G. (2024a). Does chatgpt help novice programmers write better code? results from static code analysis. *IEEE Access*, 12:114146–114160.

Haindl, P. e Weinberger, G. (2024b). Students' experiences of using chatgpt in an undergraduate programming course. *IEEE Access*, 12:43519–43530.

Hamzah, W. W., Ismail, I., Yusof, M. K., Saany, S. M., e Yacob, A. (2021). Using learning analytics to explore responses from student conversations with chatbot for education. *International Journal of Engineering Pedagogy*, 11(6):70–84.

Hien, H. T., Cuong, P.-N., Nam, L. N. H., Nhung, H. L. T. K., e Thang, L. D. (2018). Intelligent assistants in higher-education environments: the fit-ebot, a chatbot for administrative and learning support. Em *Proceedings of the 9th International Symposium on Information and Communication Technology*, páginas 69–76.

Hobert, S. (2019). Say hello to 'coding tutor'! design and evaluation of a chatbot-based learning system supporting students to learn to program.

Ismail, M. e Ade-Ibijola, A. (2019). Lecturer's apprentice: A chatbot for assisting novice programmers. Em *2019 international multidisciplinary information technology and engineering conference (IMITEC)*, páginas 1–8. IEEE.

Jury, B., Lorusso, A., Leinonen, J., Denny, P., e Luxton-Reilly, A. (2024). Evaluating llm-generated worked examples in an introductory programming course. Em *Proceedings of the 26th Australasian computing education conference*, páginas 77–86.

Kasinathan, V., Mustapha, A., Siow, S., e Hopman, M. (2018). Tictad: A chatterbot for learning visual c# programming based on expert system,". *Indonesian Journal of Electrical Engineering and Computer Science*, 11(2):740–746.

Keuning, H., Jeuring, J., e Heeren, B. (2018). A systematic literature review of automated feedback generation for programming exercises. *ACM Transactions on Computing Education (TOCE)*, 19(1):1–43.

Kiesler, N., Lohr, D., e Keuning, H. (2023). Exploring the potential of large language models to generate formative programming feedback. Em *2023 IEEE Frontiers in Education Conference (FIE)*, páginas 1–5. IEEE.

Kitchenham, B. e Charters, S. (2007). Guidelines for performing systematic literature reviews in software engineering.

Kitchenham, B., Madeyski, L., e Budgen, D. (2022). Segress: Software engineering guidelines for reporting secondary studies. *IEEE Transactions on Software Engineering*, 49(3):1273–1298.

Kosar, T., Ostojic, D., Liu, Y. D., e Mernik, M. (2024). Computer science education in chatgpt era: Experiences from an experiment in a programming course for novice programmers. *Mathematics*, 12(6):629.

Kuhrmann, M., Fernández, D. M., e Daneva, M. (2017). On the pragmatic design of literature studies in software engineering: an experience-based guideline. *Empirical software engineering*, 22:2852–2891.

Kumar, Y., Manikandan, A., Li, J. J., e Morreale, P. (2024). Preliminary results from integrating chatbots and low-code ai in computer science coursework. Em *2024 IEEE Integrated STEM Education Conference (ISEC)*, páginas 1–4. IEEE.

Kuo, Y.-C. e Chen, Y.-A. (2023). The impact of chatbots using concept maps on correction outcomes–a case study of programming courses. *Education and Information Technologies*, 28(7):7899–7925.

Lane, H. C. e VanLehn, K. (2003). Coached program planning: Dialogue-based support for novice program design. Em *Proceedings of the 34th SIGCSE technical symposium on Computer science education*, páginas 148–152.

Lee, J. S., Liu, F., e Cai, T. (2024). Code debugging with llm-generated explanations of programming error messages. Em *2024 IEEE 13th International Conference on Engineering Education (ICEED)*, páginas 1–5. IEEE.

Lepp, M. e Kaimre, J. (2025). Does generative ai help in learning programming: Students' perceptions, reported use and relation to performance. *Computers in Human Behavior Reports*, 18:100642.

Lin, Y.-H. (2022). Chatbot script design for programming language learning. Em *Proceedings of the 5th IEEE Eurasian Conference on Educational Innovation (ECEI)*. IEEE.

Liu, S., Yu, Z., Huang, F., Bulbulia, Y., Bergen, A., e Liut, M. (2024). Can small language models with retrieval-augmented generation replace large language models when learning computer science? Em *Proceedings of the ACM Conference on Com- puting Education*. ACM.

López-Pernas, S., Gordillo, A., Barra, E., e Quemada, J. (2019). Examining the use of an educational escape room for teaching programming in a higher education setting. *IEEE Access*, 7:31723–31737.

Luxton-Reilly, A. (2016). Learning to program is easy. Em *Proceedings of the 2016 ACM Conference on Innovation and Technology in Computer Science Education*, páginas 284–289.

Mageira, K., Pittou, D., Papasalouros, A., Kotis, K., Zangogianni, P., e Daradoumis, A. (2022). Educational ai chatbots for content and language integrated learning. *Applied Sciences*, 12(7):3239.

Mitchell, C. M., Boyer, K. E., e Lester, J. C. (2013). When to intervene: Toward a markov decision process dialogue policy for computer science tutoring. Em *The First Workshop on AI-supported Education for Computer Science (AIEDCS 2013)*, página 40. Citeseer.

Modran, H., Ursutiu, D., Samoila, C., e Gherman Dolhăscu, E.-C. (2024). Developing a gpt chatbot model for students programming education. Em *Advances in Intelligent Systems and Computing*. Springer.

Nguyen, H. D., Tran, T.-V., Pham, X.-T., Huynh, A. T., Pham, V. T., e Nguyen, D. (2022). Design intelligent educational chatbot for information retrieval based on integrated knowledge bases. *IAENG International Journal of Computer Science*, 49(2):531–541.

Okonkwo, C. W. e Ade-Ibijola, A. (2020). Python-bot: A chatbot for teaching python programming. *Engineering Letters*, 29(1).

Okonkwo, C. W. e Ade-Ibijola, A. (2021). Chatbots applications in education: A systematic review. *Computers and Education: Artificial Intelligence*, 2:100033.

Okonkwo, C. W. e Ade-Ibijola, A. (2022). Revision-bot: A chatbot for studying past questions in introductory programming. *IAENG International Journal of Computer Science*, 49(3).

Palahan, S. (2025). Pythonpal: Enhancing online programming education through chatbot-driven personalized feedback. *IEEE Transactions on Learning Technologies*.

Penney, J., Pimentel, J. F., Steinmacher, I., e Gerosa, M. A. (2023). Anticipating user needs: Insights from design fiction on conversational agents for computational thinking. Em *International Workshop on Chatbot Research and Design*, páginas 204–219. Springer.

Petersen, K., Vakkalanka, S., e Kuzniarz, L. (2015). Guidelines for conducting systematic mapping studies in software engineering: An update. *Information and Software Technology*, 64:1–18.

Pirzado, F. A., Ahmed, A., Mendoza-Urdiales, R. A., e Terashima-Marin, H. (2024). Navigating the pitfalls: Analyzing the behavior of llms as a coding assistant for computer science students-a systematic review of the literature. *IEEE Access*.

Prather, J., Reeves, B. N., Leinonen, J., MacNeil, S., Randrianasolo, A. S., Becker, B. A., Kimmel, B., Wright, J., e Briggs, B. (2024). The widening gap: The benefits and harms of generative ai for novice programmers. Em *Proceedings of the ACM Conference on International Computing Education Research V.1 (ICER '24)*, páginas 469–486, Melbourne, VIC, Australia. ACM.

Renkl, A. (2014). Toward an instructionally oriented theory of example-based learning. *Cognitive science*, 38(1):1–37.

Robins, A. V. (2019). 12 novice programmers and introductory programming. *The Cambridge handbook of computing education research*, página 327.

Ruan, S., Jiang, L., Xu, J., Tham, B. J.-K., Qiu, Z., Zhu, Y., Murnane, E. L., Brunskill, E., e Landay, J. A. (2019). Quizbot: A dialogue-based adaptive learning system for factual knowledge. Em *Proceedings of the 2019 CHI Conference on Human Factors in Computing Systems*, páginas 1–13.

Scholl, A., Schiffner, D., e Kiesler, N. (2025). Students' use of chatgpt in an introductory programming course: A deep dive into chat protocols and the student perspective. *eleed*, Issue 16:1–20.

Smutny, P. e Schreiberova, P. (2020). Chatbots for learning: A review of educational chatbots for the facebook messenger. *Computers & Education*, 151:103862.

Sweller, J., Ayres, P., e Kalyuga, S. (2011). *Cognitive Load Theory*. Springer.

Sweller, J., Van Merrienboer, J. J., e Paas, F. G. (1998). Cognitive architecture and instructional design. *Educational psychology review*, páginas 251–296.

Tayeb, A., Alahmadi, M., Tajik, E., e Haiduc, S. (2024). Investigating developers' preferences for learning and issue resolution resources in the chatgpt era. *arXiv preprint arXiv:2410.08411*.

Troussas, C., Krouska, A., Papakostas, C., Mylonas, P., e Sgouropoulou, C. (2024). Assessing the impact of integrating chatgpt as an advice generator in educational software. Em *2024 9th South-East Europe Design Automation, Computer Engineering, Computer Networks and Social Media Conference (SEEDA-CECNSM)*, páginas 127–133. IEEE.

Vadaparty, A., Geng, F., Smith IV, D. H., Benario, J. G., Zingaro, D., e Porter, L. (2025). Achievement goals in cs1-llm. Em *Proceedings of the 27th Australasian Computing Education Conference*, páginas 144–153.

Van Merrienboer, J. J. e Sweller, J. (2005). Cognitive load theory and complex learning: Recent developments and future directions. *Educational psychology review*, 17:147–177.

Verleger, M. e Pembridge, J. (2018). A pilot study integrating an ai-driven chatbot in an introductory programming course. Em *2018 IEEE frontiers in education conference (FIE)*, páginas 1–4. IEEE.

Vintila, F. (2024). Avert (authorship verification and evaluation through responsive testing): an llm-based procedure that interactively verifies code authorship and evaluates student understanding. Em *Proceedings of the 21st International Conference on Information Technology Based Higher Education and Training (ITHET)*. IEEE.

Wei, S., Luo, Y., Chen, S., Huang, T., e Xiang, Y. (2023). Deep research and analysis of chatgpt based on multiple testing experiments. Em *2023 International Conference on Cyber-Enabled Distributed Computing and Knowledge Discovery (CyberC)*, páginas 123–131. IEEE.

Wijaya, O. C. e Purwarianti, A. (2024). An interactive question-answering system using large language model and retrieval-augmented generation in an intelligent tutoring system on the programming domain. Em *Proceedings of the 2024 International Conference on Advanced Informatics: Concept, Theory and Application (ICAICTA)*. IEEE.

Winkler, R., Hobert, S., Salovaara, A., Söllner, M., e Leimeister, J. M. (2020). Sara, the lecturer: Improving learning in online education with a scaffolding-based conversational agent. Em *Proceedings of the 2020 CHI conference on human factors in computing systems*, páginas 1–14.

Xiao, R., Hou, X., Kumar, H., Moore, S., Stamper, J., e Liut, M. (2024). A preliminary analysis of students' help requests with an llm-powered chatbot when completing cs1 assignments. Em *Proceedings of the 8th Educational Data Mining in Computer Science Education Workshop (CSEDM'24)*, Atlanta, GA. CEUR-WS, CEUR Workshop Proceedings.

Xue, Y., Chen, H., Bai, G. R., Tairas, R., e Huang, Y. (2024). Does chatgpt help with introductory programming? an experiment of students using chatgpt in cs1. Em *Proceedings of the 46th International Conference on Software Engineering: Software Engineering Education and Training*, páginas 331–341.

Zabala, E. e Narman, H. S. (2024). Development and evaluation of an ai-enhanced python programming education system. Em *Proceedings of the 15th IEEE Annual Ubiquitous Computing, Electronics and Mobile Communication Conference*. IEEE.

Zhang, H., Kitchenham, B., e Pfahl, D. (2010). Software process simulation modeling: an extended systematic review. Em *International Conference on Software Process*, páginas 309–320. Springer.